\documentstyle[12pt]{article}
\begin{document}
\newcommand{\gtrsim}{ \mathop{}_{\textstyle \sim}^{\textstyle >} }
\newcommand{\lesssim}{ \mathop{}_{\textstyle \sim}^{\textstyle <} }
\newcommand{\eV}{~\mbox{eV}}
\newcommand{\keV}{~\mbox{keV}}
\newcommand{\MeV}{~\mbox{MeV}}
\newcommand{\GeV}{~\mbox{GeV}}
\newcommand{\TeV}{~\mbox{TeV}}
\begin{flushright}
UT-853\\
RESCEU-15/99
\end{flushright}
\vskip 2cm
\begin{center}
{\Large \bf Leptogenesis in Inflaton Decay \\}
\vskip 1cm
{\large T. Asaka$^1$, K. Hamaguchi$^1$, 
M. Kawasaki$^2$ and T. Yanagida$^{1,2}$} \\
\vskip 0.5cm

{\em $^1$Department of Physics,  University of
  Tokyo, Tokyo 113-0033, Japan}
\vskip 0.2cm
{\em $^2$Research Center for the Early Universe,
  University of Tokyo, Tokyo 113-0033, Japan}
\vskip 0.5cm
{(June 15, 1999)}
\end{center}
\vskip 2cm
\begin{abstract}
    We study a leptogenesis via decays of heavy Majorana neutrinos 
    produced non-thermally in inflaton decays. We find
    that this scenario is fully consistent with existing 
    supersymmetric inflation
    models such as for topological or for hybrid inflation and the
    Froggatt-Nielsen mechanism generating hierarchies in quark and
    lepton mass matrices.
    The reheating temperature $T_R$ of inflation may be taken as
    low as $T_R \simeq 10^8$ GeV to avoid the cosmological gravitino 
    problem.
\end{abstract}
\thispagestyle{empty} \setcounter{page}{0} \newpage
\setcounter{page}{1}
\baselineskip 0.65cm
\section{Introduction}
\label{sec:Introduction}
%
Primordial lepton asymmetry is converted to baryon asymmetry
\cite{Fukugita-Yanagida} in the early universe 
through the ``sphaleron'' effects of the electroweak gauge
theory \cite{Kuzmin-Rubakov-Shaposhnikov} if it is produced before the 
electroweak phase transition. 
Therefore, lepton-number violation at high energies
may be an important ingredient for creating the baryon
(matter-antimatter) asymmetry in the present universe.
A generic low-energy prediction of the leptogenesis is
the presence of effective operators for the lepton-number
violation such as
\begin{eqnarray}
    \label{LV-O}
    {\cal O} = \frac{ f_{ij} }{ M } 
    l_i l_j H H,
\end{eqnarray}
where $M$ is the scale of lepton-number violation,
$l_i$ ($i$=1,2,3) lepton doublets and $H$ the Higgs scalar field
in the standard model.

The above operators induce small neutrino masses
\cite{See-Saw}
in the true vacuum $\langle H \rangle \simeq$ 246/$\sqrt{2}$ GeV.
There is now a convincing experimental evidence
\cite{Super-K}
that neutrinos have indeed small masses $m_\nu$ of order
0.01--0.1 eV.
Thus, the leptogenesis scenario \cite{Fukugita-Yanagida}
seems to be the most plausible mechanism for creating the cosmological 
baryon asymmetry.

The lepton-number violating operators Eq. (\ref{LV-O})
arise from the exchange of heavy Majorana neutrinos
$N_i$ ($i$=1,2,3).
Decays of these heavy neutrinos
produce very naturally lepton asymmetry
if $C$ and $CP$ are not conserved \cite{Fukugita-Yanagida}.
There have been, so far, proposed and investigated
various scenarios for the leptogenesis depending on 
production mechanisms of the heavy Majorana neutrinos
$N_i$ \cite{Fukugita-Yanagida,LG-thermal,LG-infdecay,LG-osc}.

In this letter we discuss a leptogenesis scenario where
$N_i$ are produced non-thermally in inflaton decays 
\cite{LG-infdecay}.
We find that this scenario is fully consistent with existing
inflation models such as for topological \cite{Izawa-Kawasaki-Yanagida}
or for hybrid inflation \cite{Hybrid-Inf-Model},
and the Froggatt-Nielsen (FN) mechanism 
\cite{Froggatt-Nielsen} generating
hierarchies in the quark and lepton mass matrices.
The reheating temperature $T_R$ of inflation may be taken as low as 
$T_R \simeq 10^8$ GeV avoiding the cosmological gravitino problem
\cite{Gravitino-Prob}.

We assume supersymmetry (SUSY) throughout this letter.
In SUSY models there is an interesting leptogenesis mechanism
\cite{Murayama-Yanagida}
of the Affleck-Dine type 
whose detailed analysis will be given elsewhere.
\section{Lepton asymmetry in the decays of heavy Majorana neutrinos}
\label{sec:LA}
%
Decays of the heavy neutrinos $N$ into leptons $l$
and Higgs $H_u$ doublets violate lepton-number conservation
since they possess two decay channels,%
\footnote{
Here $N$, $l (\overline{l})$ and $H_u (\overline{H_u})$ denote
fermionic or bosonic components of corresponding supermultiplets.  }
\begin{eqnarray}
    && N \rightarrow H_u + l \nonumber \\
    && N \rightarrow \overline{H_u} + \overline{l}.
\end{eqnarray}
We consider only the $N_1$ decay, provided the mass $M_1$ 
of $N_1$ is smaller than the others
($M_1 \ll M_2, M_3$).
Interference between amplitudes of the tree-level and 
one-loop diagrams yields a lepton asymmetry 
\cite{Covi,Buchmuller-Plumacher}
\begin{eqnarray}
    \epsilon_1 
    &\equiv&
    \frac{ \Gamma (N_1 \rightarrow H_u + l )
         - \Gamma (N_1 \rightarrow \overline{H_u} + \overline{l} ) }
         { \Gamma (N_1 \rightarrow H_u + l )
         + \Gamma (N_1 \rightarrow \overline{H_u} + \overline{l} ) }
    \\
    &=&
    - 
    \frac{ 3 }{ 16 \pi \left( h_\nu h_\nu^{\dagger} \right)_{11} }
    \left[ 
        \mbox{Im} \left( h_\nu h_\nu^{\dagger} \right)_{13}^2 
        \frac{ M_1 }{ M_3 }
        +
        \mbox{Im} \left( h_\nu h_\nu^{\dagger} \right)_{12}^2 
        \frac{ M_1 }{ M_2 }
    \right]
    .
    \label{Ep1}
\end{eqnarray}
Here we have taken a basis where the mass matrix for $N_i$ is 
diagonal and the Yukawa coupling constants $(h_\nu)_{ij}$ are defined
in the superpotential as $W = (h_\nu)_{ij} N_i l_j H_u$.
We have included both of one-loop vertex and self-energy corrections
\cite{Buchmuller-Plumacher}.

The FN model considered in the next section suggests 
$\left( h_\nu h_\nu^{\dagger} \right)_{13}^2 / M_3
\simeq \left( h_\nu h_\nu^{\dagger} \right)_{12}^2 /M_2$,
and hence we rewrite the lepton asymmetry parameter $\epsilon_1$
in Eq. (\ref{Ep1}) by using an effective $CP$-violating
phase $\delta_{\rm eff}$ as
\begin{eqnarray}
    \epsilon_1 \simeq
    \frac{ 3 \delta_{\rm eff} }
         { 16 \pi \left( h_\nu h_\nu^{\dagger} \right)_{11} }
    \left| \left( h_\nu h_\nu^\dagger \right)^2_{13} \right|
    \frac{ M_1 }{ M_3 }.
\end{eqnarray}
Assuming $\left | (h_\nu)_{i3} \right| > \left| (h_\nu)_{i2}
\right| \gg \left|(h_\nu)_{i1} \right| (i = 1,3)$ we obtain
\begin{eqnarray}
    \epsilon_1 
    &\simeq&
    \frac{ 3 \delta_{\rm eff} }{ 16 \pi }
    \left| (h_\nu)_{33}^2 \right|
    \frac{ M_1 }{ M_3 } \nonumber \\
    &\simeq&
    \frac{ 3 \delta_{\rm eff} }{ 16 \pi }
    \frac{ m_{\nu_3} M_1 }{ \langle H_u \rangle^2 }.
\end{eqnarray}
Here, we have used the see-saw formula \cite{See-Saw}
[see Eq. (\ref{mnu})]
\begin{eqnarray}
    m_{\nu_3} \simeq
    \frac{ \left| ( h_{\nu} )_{33}^2 \right|
           \langle H_u \rangle^2 }
    {M_3}.
\end{eqnarray}
Taking the maximal $CP$ violating phase $| \delta_{\rm eff} | \simeq 1$,
$m_{\nu_3} \simeq 3 \times 10^{-2}$ eV suggested%
\footnote{
The atmospheric neutrino oscillation observed in the Superkamiokande
experiments indicates the neutrino mass squared difference
$m_{\nu_3}^2 - m_{\nu_2}^2 \simeq (0.5$--$6) \times 10^{-3}$ eV$^2$
\cite{Super-K}.
Assuming mass hierarchy in the neutrino sector,
$m_{\nu_3} \gg m_{\nu_2}$, we obtain 
$m_{\nu_3} \simeq (2$--$8) \times 10^{-2}$ eV.
}
from the Superkamiokande experiments \cite{Super-K}, and
$\langle H_u \rangle \simeq$ 174/$\sqrt{2}$ GeV,%
\footnote{
   In the SUSY standard model $\langle H_u \rangle = \sin \beta ~
   \langle H \rangle$.
   We assume, here, $\tan \beta \equiv \langle H_u \rangle 
   / \langle H_d \rangle \simeq {\cal O}$(1),
   where $H_u$ and $H_d$ are Higgs supermultiplets couple to
   up-type and down-type quarks, respectively.
   If one takes $\tan \beta \simeq 50$ the lepton asymmetry
   is reduced only by factor 2.
}
we get \cite{Buchmuller-Yanagida}
\begin{eqnarray}
    \epsilon_1 \simeq -
    10^{-6} \left( \frac{ M_1 }{ 10^{10}~\mbox{GeV} } \right).
\end{eqnarray}

Let us now consider the production of heavy neutrino $N_1$ 
through inflaton $\varphi$ decays, which leads to a 
constraint on the inflaton mass $m_\varphi$ as
\begin{eqnarray}
    \label{const-mphi0}
    m_\varphi > 2 M_1.
\end{eqnarray}
We consider the case where $M_1 \gtrsim 100 T_R$ 
($T_R$ is the reheating temperature of the inflation).%
\footnote{
Inflation models we discuss in this letter satisfy this condition.
}
In this case the Majorana neutrino $N_1$ is always out of thermal 
equilibrium even if it has $\cal O$(1) Yukawa coupling 
to $H_u$ and $l_i$,
and the $N_1$ behaves like frozen-out,
relativistic particle with the energy $E_{N1} \simeq m_\varphi /2$.
As we will see in the next section,
the $N_1$ decays immediately after produced by the inflaton decays
and hence we obtain lepton-to-entropy ratio \cite{LG-infdecay}
\begin{eqnarray}
    \frac{ n_L }{s} 
    &\simeq&
    \frac{ 3 }{ 2 } 
    \epsilon_1 B_r
    \frac{ T_R }{ m_\varphi } 
    \nonumber \\
    &\simeq& -
    10^{-6} B_r
    \left( \frac{ T_R }{ 10^{10} ~\mbox{GeV} } \right)
    \left( \frac{ M_1 }{ m_\varphi } \right),
    \label{LA}
\end{eqnarray}
where $B_r$ is the branching ratio of the inflaton decay into 
$N_1$ channel.
We impose $T_R \lesssim 10^{8}$ GeV which is required
to suppress sufficiently the density of the 
gravitino \cite{Gravitino-Prob}.
Notice that
the reheating temperature $T_R$ is bounded from below,
$T_R \gtrsim 10^6$ GeV, otherwise the produced lepton asymmetry 
is too small as $n_L / s < 10^{-10}$.

The lepton asymmetry in Eq. (\ref{LA}) is converted to the baryon
asymmetry through the ``sphaleron'' effects which is given by 
\begin{eqnarray}
    \frac{ n_B }{ s } \simeq a \frac{ n_L }{ s },
\end{eqnarray}
with $a \simeq - 8/23$ \cite{Khlebnikov-Shaposhnikov} in the minimal 
SUSY standard model. 
To explain the observed baryon asymmetry 
\begin{eqnarray}
    \frac{ n_B }{ s } 
    \simeq ( 0.1 \mbox{--} 1 ) \times 10^{-10},
\end{eqnarray}
we get a constraint 
\begin{eqnarray}
    \label{const-mphi}
    2 <
    \frac{ m_\varphi }{ M_1 } 
    \lesssim 100 B_r
    \lesssim 100.
\end{eqnarray}

\section{Froggatt-Nielsen model for neutrino masses and mixing
angles}
\label{sec:FNmodel}
%
The Froggatt-Nielsen (FN) mechanism \cite{Froggatt-Nielsen}
is the most attractive framework
for explaining the observed hierarchies in the quark and charged 
lepton mass matrices, which is based on a 
broken $U(1)_F$ symmetry.
A gauge singlet field $\Phi$ carrying the FN charge 
$Q_\Phi = -1$ is assumed to have a vacuum-expectation value
$\langle \Phi \rangle$ and then the Yukawa couplings
of Higgs supermultiplets
arise from nonrenormalizable interactions of $\Phi$ as
\begin{eqnarray}
    W = 
    g_{ij} 
    \left( \frac{ \Phi }{ M_G } \right)^{Q_i + Q_j }
    \Psi_i \Psi_j H_{u(d)},
\end{eqnarray}
where $Q_i$ are the $U(1)_F$ charges of various supermultiplets
$\Psi_i$,
$g_{ij}$ ${\cal O}(1)$ coupling constants,
and the gravitational scale $M_G \simeq 2.4 \times 10^{18}$ GeV.
The mass hierarchies for quarks and charged leptons are well 
explained in terms of their FN charges with 
$\epsilon \equiv \langle \Phi \rangle /M_G \simeq 1/17$ 
\cite{Sato-Yanagida}.

We adopt the above mechanism to the neutrino sector.
Possible FN charges for various supermultiplets are shown in 
Table \ref{tab:FNcharges}.
\begin{table}[t]
    \begin{center}
    \begin{tabular}{| c | c c c | c c c  | c c c | }
        \hline
        $\Psi_i$ & $l_3$ & $l_2$ & $l_1$ 
                 & $e_3^c$ & $e_2^c$ & $e_1^c$ 
                 & $N_3$ & $N_2$ & $N_1$ \\
        \hline
        $Q_i$    & $a$   & $a$   & $a+1$
                 & 0     & 1     & 2
                 & b     & c     & d \\
        \hline
    \end{tabular}
    \caption{The FN charges of various supermultiplets. 
    We assume $a=0$ or 1 and $b \le c < d$ in the text.}
    \label{tab:FNcharges}
    \end{center}
\end{table}
The charges for $e_i^c$ are taken to be the same as those of 
up-type quarks assuming that they belong to the same {\bf 10}'s 
in the $SU(5)$ grand unified theory.
The charge $a$ of the $l_3$ may be 0 or 1
(see Ref. \cite{Sato-Yanagida} for details).
The assignment of the FN charges for the lepton doublets
yields a mass matrix for neutrinos
\cite{Sato-Yanagida,Buchmuller-Yanagida} as
\begin{eqnarray}
    \label{mnu}
    (m_{\nu})_{ij}
    \simeq
    \epsilon^{2 a}
    \left( 
        \begin{array}{ c c c}
            \epsilon^2 & \epsilon & \epsilon \\
            \epsilon   & 1        & 1 \\
            \epsilon   & 1        & 1 
        \end{array}
    \right)
    \frac{ \langle H_u \rangle^2 }{ M_0 } ~.
\end{eqnarray}
Here, Majorana masses $M_i$ of the heavy neutrinos $N_i$ 
are given by $M_i = \epsilon^{2 Q_i} M_0$ with $Q_i$ being
the FN charges of $N_i$.
This mass matrix leads to a large $\nu_\mu$--$\nu_\tau$
mixing angle, which is consistent with the atmospheric neutrino
oscillation \cite{Sato-Yanagida}.
Using $m_{\nu_3} \simeq 3 \times 10^{-2}$ eV we may derive
\begin{eqnarray}
    M_0 \simeq \epsilon^{2a} \times 10^{15} ~\mbox{GeV} ~
    \simeq
    \left\{
        \begin{array}{ll}
            10^{15} ~\mbox{GeV} ~~~~~~&\mbox{for}~~ a = 0\\
            10^{13} ~\mbox{GeV} ~~~~~~&\mbox{for}~~ a = 1
        \end{array}
    \right.~.
\end{eqnarray}

Hereafter,
we assume $a=0$ and $M_0 \simeq 10^{15}$ GeV
for simplicity. 
Notice that the final result depends only on $B_r$,
$T_R$ and $M_1/m_\varphi$ as seen in Eq. (\ref{LA}),
and hence the FN charge $a$ itself is irrelevant to the present analysis.
For $d=1$ and $d=2$ we obtain $M_1\simeq 10^{13}$ GeV 
and $10^{10}$ GeV, respectively.%
\footnote{
For the case of $a=1$
we obtain $M_1 \simeq 10^{10}$ GeV for $d=1$.
}
The partial decay rate $\Gamma ( N_1 \rightarrow H_u l )$ is given by
\begin{eqnarray}
    \Gamma ( N_1 \rightarrow H_u l )
    \simeq \frac{ 1 }{ 8 \pi } 
    \left| \epsilon \right|^{2(a+d)} M_1.
\end{eqnarray}

For the case of $M_1 \simeq 10^{10}$ GeV we see from
Eq. (\ref{const-mphi}) that 
the inflaton mass should lie in a range of 
$10^{10}$ GeV $\lesssim m_\varphi \lesssim 10^{12}$ GeV 
to explain the present observed baryon asymmetry.
We will find in the following section that 
a topological inflation model \cite{Izawa-Kawasaki-Yanagida}
proposed recently satisfies
this condition and $T_R \simeq 10^{8}$ GeV.
The total decay rate of the $N_1$ is $\Gamma_{N_1}  \simeq
(1 / 4 \pi) \left| \epsilon \right|^4 M_1 \simeq 10^4$ GeV which is
much larger than the inflaton decay rate
$\Gamma_\varphi \sim T_R^2 / M_G \simeq 10^{-2}$ GeV,
and hence the $N_1$ decays immediately after produced by the inflaton
decay.
\footnote{
Perturbative calculation in the present analysis is justified,
since $\Gamma_{N_1} \ll M_1$.
}

For the case of $M_1 \simeq 10^{13}$ GeV we find from
Eq. (\ref{const-mphi}) that the inflaton mass of
$10^{13}$ GeV $\lesssim m_\varphi \lesssim 10^{15}$ GeV is required,
which nicely fits in a SUSY hybrid inflation model \cite{Hybrid-Inf-Model}
as seen in the next section.
We find that $\Gamma_{N_1} \gg \Gamma_\varphi$ in this case 
and our approximation used in deriving Eq. (\ref{LA}) is also justified.

\section{Leptogenesis in inflaton decays}
\label{sec:LG}

We are now at the point to discuss the leptogenesis in inflaton
decays.  We first consider the case of $M_1 \simeq 10^{10}$ GeV and
a topological inflation model with 
the inflaton mass of $10^{10}$ GeV $\lesssim m_\varphi \lesssim
10^{12}$ 
GeV.
We adopt a SUSY topological inflation which has been proposed in
Ref.~\cite{Izawa-Kawasaki-Yanagida}.  The model has the following
superpotential $W$ and K\"ahler potential $K$;
\begin{eqnarray}
    W & = & v^2 \chi ( 1 - g \phi^2/M_G^2), \\
    K & = & |\chi|^2 + |\phi|^2
    + \frac{1}{M_G^2} \left( k_{1}|\chi|^2|\phi|^2 
    -\frac{k_{2}}{4}|\chi|^4 \right)  + \cdots ,
\end{eqnarray}
where $v$ is the energy scale of the inflation, $g$, $k_{1}$ and
$k_{2}$ are constants of order unity,
and the ellipsis denotes higher order terms.
  Here, we impose $U(1)_R \times
Z_{2}$ symmetry and omit higher-order terms for simplicity.  
We assume $\chi (\theta) \rightarrow e^{-2i\alpha} \chi(\theta
e^{i\alpha})$ and $\phi(\theta) \rightarrow \phi(\theta
e^{i\alpha})$ under the  $U(1)_{R}$,
and  $\chi$ is even and $\phi$ is odd under the $Z_2$.  
This discrete $Z_2$ symmetry is an essential ingredient for
the topological inflation \cite{Topological-Inf}.

The potential of scalar components of the supermultiplets
$\chi$ and $\phi$ is obtained in the standard manner,
which yields SUSY vacua
\begin{eqnarray}
    \langle \chi \rangle = 0,
    \quad
    \langle \phi \rangle = \pm
    \frac{M_G}{\sqrt{g}} \equiv \pm \eta,
    \label{eq:eta-g}
\end{eqnarray}
in which the potential energy vanishes. Here, the scalar components of
the supermultiplets are denoted by the same symbols as the corresponding
supermultiplets.  

A topological inflation~\cite{Topological-Inf} occurs if the 
vacuum-expectation value $\langle \phi \rangle$ is of order of the
gravitational scale $M_G$.  The critical value $\eta_{\rm cr}$ of $\langle
\phi \rangle$ for which the topological inflation occurs was
investigated in Refs.\cite{Sakai,Cho}. We adopt the result in
Ref.~\cite{Sakai}, which gives
\begin{equation}
     \eta_{\rm cr} \simeq 1.7 M_G.
\end{equation}
Thus, for topological inflation to take place, $\eta$ should be
larger than $\eta_{\rm cr}$.

The potential for the region $|\chi|$, $|\phi| \ll M_G$
is written approximately as
\begin{equation}
    \label{eff-pot}
    V \simeq v^4|1 - g\phi^{2}/M_G^2|^{2} + (1 - k_1)v^4 |\phi|^2/M_G^2
    + k_2v^4 |\chi|^2/M_G^2.
\end{equation}
Since the $\chi$ field quickly settles down to the origin for the case 
$k_2 \gtrsim 1$, we set $\chi =0$ in Eq. (\ref{eff-pot}) 
assuming $k_2 \gtrsim 1$.  
For $g > 0$ and $k_1 < 1$, we identify the inflaton field
$\varphi/\sqrt{2}$ with the real part of the field $\phi$ since
the imaginary part of $\phi$ has a positive mass larger than the size
of the negative mass of $\varphi$ near the origin $\phi \simeq 0$. 
Then, we obtain a potential for the
inflaton as
\begin{equation}
    \label{eff-pot2}
    V(\varphi) \simeq v^4 - \frac{\kappa}{2M_G^2}v^4\varphi^2,
\end{equation}
where $\kappa \equiv 2g + k_{1} - 1$.
The inflaton has a mass $m_{\varphi}$ 
\begin{equation}
    \label{inflaton-mass}
    m_{\varphi} \simeq 2|\sqrt{g}v^2/M_G| = 2v^2/\eta,
\end{equation}
in the true vacuum Eq. (\ref{eq:eta-g}).

The slow-roll condition for the inflation is satisfied for $0< \kappa
< 1$ and $\varphi < \varphi_f$ where $\varphi_f$ is of order of $M_G$,
which provides the value of $\varphi$ at the end of inflation. 
The scale factor of the universe increases by a factor of $e^N$ when 
the inflaton $\varphi$ rolls slowly down the potential from 
$\varphi_N$ to $\varphi_f$.  The $e$-fold number $N$ is given by
\begin{eqnarray}
    N \simeq \int^{\varphi_{N}}_{\varphi_f}
      d\varphi \frac{V}{V'M_G^2}
      \simeq {1 \over \kappa}\ln{\varphi_f \over \varphi_N}.
    \label{N-efold}
\end{eqnarray}

The amplitude of primordial density fluctuations $\delta \rho/\rho$ 
due to the inflation is written as 
\begin{equation}
    \label{density}
    \frac{\delta\rho}{\rho} \simeq \frac{1}{5\sqrt{3}\pi M_G^3}
    \frac{V^{3/2}(\varphi_{N})}{|V'(\varphi_{N})|}
    \simeq \frac{1}{5\sqrt{3}\pi} \frac{v^{2}}{\kappa\varphi_{N}M_G}.
\end{equation}
This  should be normalized to the
data on anisotropies of the cosmic microwave background radiation
(CMBR) 
by the Cosmic Background Explorer (COBE) 
satellite \cite{COBE}.  Since the $e$-fold number $N$
corresponding to the COBE scale is about $60$, the COBE
normalization gives
\begin{equation}
    \label{COBE-norm}
    \frac{V^{3/2}(\varphi_{60})}{M_G^3|V'(\varphi_{60})|}
    \simeq 5.3\times 10^{-4}.
\end{equation}
In this model the spectrum index $n_s$ is given by
(for details see Ref. \cite{Izawa-Kawasaki-Yanagida})
\begin{equation}
    \label{new-index}
    n_s \simeq 1 - 2 \kappa.
\end{equation}
Since the COBE data implies $n_{s}$ as $n_{s} = 1.0 \pm 0.2$
\cite{COBE}, we should 
take $\kappa \lesssim 0.1$. Along with Eqs. (\ref{N-efold}) and
(\ref{COBE-norm}), we obtain
\begin{eqnarray}
    v  \simeq  2.3 \times 10^{-2} \sqrt{\kappa \varphi_f M_G} ~
    e^{- \frac{\kappa N}{2}},
\end{eqnarray}
which leads to
\begin{equation}
    m_{\varphi}  \simeq  3.8\times 10^{11}{\rm GeV},
\end{equation}
for $\kappa \simeq 0.1$, $\eta \simeq \eta_{\rm cr}$ and 
$\varphi_{f}\simeq M_G$. %
\footnote{
If one takes $\kappa \lesssim 0.07$, one gets
$m_\phi \gtrsim 10^{12}$ GeV.
}

The inflaton decay into $2N_{1}$ occurs through 
nonrenormalizable interactions
in the K\"ahler potential such as
\begin{equation}
    \label{K_TI}
    K = \sum_i  C_i  |\phi|^2|\psi_i|^2/M_G^2,
\end{equation}
where $\psi_i$ denote supermultiplets for SUSY standard-model 
particles including $N_1$,
and $C_i$ are coupling constants of order unity. With these
interactions the decay rate $\Gamma_{\varphi}$ of the inflaton is
estimated as
\begin{equation}
    \Gamma_{\varphi} \simeq 
    \sum_i 
    \frac{ C_i^2 \eta^2  m_{\varphi}^3 }{8 \pi M_G^4 }, 
\end{equation}
which yields the reheating temperature $T_R$ given by
\begin{equation}
    \label{reheat-temp}
    T_R \simeq 0.092 C\eta
    (m_{\varphi}/M_G)^{3/2} \simeq 10^{8} {\rm GeV},
\end{equation}
for $C = \sqrt{ \sum_i |C_i|^2 } \simeq 4.3$, 
$\kappa \simeq 0.1$, and $\eta \simeq \eta_{\rm 
cr}$.  Therefore, this topological inflation model can provide
$T_{R}\simeq 10^{8}$~GeV and $m_{\varphi}\simeq 10^{11-12}$~GeV, 
which leads to
\begin{eqnarray}
    \frac{ n_L }{ s } 
    \simeq - 10^{-8} B_r
    \left( \frac{ M_1 }{ m_\varphi } \right) 
    \simeq - 10^{-10},
\end{eqnarray}
for $M_1 \simeq 10^{10}$ GeV and $B_r \simeq 0.1$--1.

Next, we study a hybrid inflation model~\cite{Hybrid-Inf-Model} 
which gives the inflaton mass between $10^{13}$ GeV and $10^{15}$ GeV
and $T_R \simeq 10^8$ GeV.%
\footnote{%
Leptogenesis in the hybrid inflation model has been
considered in Ref. \cite{Lazarides}, 
where the lepton asymmetry arises from the decays of the heavy 
neutrino $N_2$ in the second family.  }
The SUSY
hybrid inflation model contains two kinds of supermultiplets: one is
$\phi$ and the others are $\Psi$ and $\overline{\Psi}$.  The model is also
based on the $U(1)_R$ symmetry.  The superpotential and K\"ahler
potential are given by~\cite{Dvali,Copeland,Hybrid-Inf-Model}
\begin{eqnarray}
    \label{sp-HI}
    W & = & -\mu^{2} \phi + \lambda \phi
    \overline{\Psi}\Psi,\\
    \label{KP-HI2}
    K & = & |\phi|^{2} + |\Psi|^{2} + |\overline{\Psi}|^{2}  + \cdots ,
\end{eqnarray}
where the ellipsis denotes higher-order terms, which we neglect in the 
present analysis.  To satisfy the $D$-term flatness condition we take 
always $\Psi = \overline\Psi$ in our analysis assuming an extra $U(1)$ 
gauge symmetry.
\footnote{
One may consider the following superpotential 
instead of Eq. (\ref{sp-HI});
$W = - \mu^2 \phi + \lambda \phi \Psi^2 +
\lambda' \phi \Psi^3 / M_G + \cdots ~.$
In this case one does not need the extra $U(1)$ gauge symmetry.
The third term is introduced to erase unwanted domain walls.
}
  As shown in Ref.\cite{Hybrid-Inf-Model} 
the real part of $\phi$ is identified with the 
inflaton field $\varphi/\sqrt{2}$.  The potential is minimized at $\Psi 
= \overline{\Psi} = 0$ when $\varphi$ is larger than $\varphi_{c}= 
\sqrt{2}\mu/\sqrt{\lambda}$, and inflation occurs for 
$ \varphi_{c} < \varphi < M_G$.

Including one-loop corrections~\cite{Dvali}, the potential for the 
inflaton $\varphi$ is given by
\begin{equation}
    \label{pre-eff-pot}
    V \simeq 
    \mu^4 \left[ 1
    + \frac{\lambda^2}{8\pi^2}\ln
    \left(\frac{\varphi}{\varphi_c}\right)
    \right].
\end{equation}
The $e$-fold number $N$ is estimated as
\begin{equation}
    N \simeq  \frac{4\pi^2}{\lambda^{2}M_G^2}
     \left( \varphi_N^2 - \varphi_c^2 \right).
    \label{Ndash}
\end{equation}
For $\lambda \gtrsim 10^{-3}$ 
the scales $\mu$ and $\langle \Psi \rangle \equiv \xi$ are
determined by the COBE normalization [Eq.~(\ref{COBE-norm})] as
\begin{eqnarray}
    \label{eq:mu-N-one-loop}
    \mu & \simeq & 5.5\times 10^{15} \lambda^{1/2}{\rm GeV},\\
    \label{eq:xi-N-one-loop}
    \xi & \simeq & 5.5\times 10^{15}{\rm GeV},
\end{eqnarray}
where we have used $N\simeq 60$ at the COBE scale.
\footnote{
The spectrum index is estimated as $n_s \simeq 1$.
}
Then, the mass of 
the inflaton $m_{\varphi}$ in the true minimum 
($\langle \Psi \rangle = \langle \overline{\Psi} \rangle = \xi$
and $\langle \phi \rangle$=0)
is given by
\begin{equation}
    m_{\varphi} \simeq \sqrt{2 \lambda}\mu
    \simeq 7.7\times 10^{15} \lambda {\rm GeV},
\end{equation}
which gives $m_{\varphi}\simeq 10^{13}$--$10^{15}$~GeV for 
$\lambda \simeq 10^{-3}$--$10^{-1}$. 
We assume that the inflaton decays through nonrenormalizable
interactions in the K\"ahler potential as
\begin{eqnarray}
    \label{KP-HI}
  K = \sum_i C_i' \left| \Sigma \right|^2 \left| \psi_i \right|^2  
    /M_G^2,
\end{eqnarray}
with $\Sigma \equiv ( \Psi + \overline{\Psi} )/ \sqrt{2}$.
The reheating temperature is estimated as
\begin{equation}
    \label{reheat-temp-hybrid}
    T_R \simeq 0.092 C' \xi
    (m_{\Sigma}/M_G)^{3/2} \simeq 10^{8} {\rm GeV},
\end{equation}
for $C' = \sqrt{ \sum_i |C_i'|^2 } \simeq 40$--$0.04$.
Notice that $m_\Sigma = m_\varphi$ in the true vacuum.
The  obtained reheating temperature $T_{R}\simeq 
10^{8}$~GeV and inflaton mass $m_{\varphi}\simeq
10^{13}$--$10^{15}$~GeV
lead to the required lepton asymmetry,
$( n_L / s ) \simeq - 10^{-8} B_r ( M_1/m_\varphi ) \simeq - 10^{-10}$,
for $M_1 \simeq 10^{13}$ GeV and $B_r \simeq$ 0.01--1.

We should comment on an interesting possibility that the 
extra $U(1)$ gauge symmetry introduced to fix 
$\Psi = \overline{\Psi}$
is nothing but the $B-L$ symmetry,
which is spontaneously broken by the condensations
$\langle \Psi \rangle = \langle \overline{\Psi} \rangle = \xi$.
\footnote{
The spontaneous breakdown of the $B-L$ gauge symmetry produces 
cosmic strings. 
Astrophysical analyses on anisotropies of the CMBR 
give a constraint
$\xi \lesssim 6.3 \times 10^{15}$ GeV \cite{String}.
Thus, the present parameter region Eq. (\ref{eq:xi-N-one-loop})
may be accessible to future satellite experiments \cite{MAP}.
}
Namely, the Majorana masses $M_i$ of the heavy neutrinos $N_i$
are induced by Yukawa couplings
\footnote{
The $B-L$ charges of $N_i$, $\Psi$ and $\overline{\Psi}$ are 
$-1$, $+2$ and $-2$, respectively.
In Ref. \cite{Lazarides} 
the $B-L$ charges of $\Psi$ and $\overline{\Psi}$ are chosen
as $+1$ and $-1$ so that
one has nonrenormalizable interactions
$W = g_i N_i N_i \Psi \Psi/ M_G $.
}
\begin{eqnarray}
    \label{Yukawa}
    W = g_i N_i N_i \Psi.
\end{eqnarray}
In this model the inflaton $\varphi$ and $\Sigma$ decay
mainly through the Yukawa couplings Eq. (\ref{Yukawa}) and 
the reheating temperature is given by 
\begin{eqnarray}
    T_R \simeq 0.092 g_1 \sqrt{ m_\Sigma M_G },
\end{eqnarray}
provided $m_\varphi = m_\Sigma < 2 M_{2,3}$.
We see that the $T_R$ is too high as 
$T_R \gtrsim 10^{12}$ GeV for $M_1 \simeq 2 g_1 \langle \Psi \rangle
\simeq 10^{13}$ GeV. Thus, we consider $M_1 \simeq 10^{10}$ GeV.
In this case we get $T_R \simeq 10^{8}$ GeV for 
$m_\varphi \simeq 10^{12}$ GeV,
\footnote{
If we take $m_\varphi \simeq 10^{10}$ GeV ($> 2 M_1$),
the reheating temperature $T_R$ becomes lower as 
$T_R \simeq 10^{6}$--$10^{7}$ GeV. 
The desired lepton asymmetry $(n_L/s) \simeq - 10^{-10}$ may be 
obtained even in this case, because of $m_\varphi / M_1 \simeq 2$
and $B_r \simeq 1$ [see Eq. (\ref{LA})].
}
which yields the desired lepton asymmetry 
$(n_L / s) \simeq - 10^{-10}$, since $B_r \simeq 1$.
The small inflaton mass $m_\varphi$ or equivalently the small
coupling $\lambda \simeq 10^{-4}$ in Eq. (\ref{sp-HI}) may be 
naturally explained by a large FN charge of 
$\phi$ (e.g. $Q_\phi = 3$).
\section{Conclusions}
\label{sec:Conc}

We have studied the SUSY leptogenesis via the decay of
the heavy Majorana neutrino $N_1$'s which are produced non-thermally
in inflaton decays.  Generic argument has shown that the leptogenesis
works without overproduction of gravitinos if the reheating
temperature $T_R \simeq 10^{8}$~GeV and the inflaton mass $m_{\varphi}$
satisfies $2 M_1 < m_{\varphi} \lesssim 100 M_{1}$. With use of the
Froggatt-Nielsen model the mass of the heavy Majorana neutrino $N_1$
in the first family is easily chosen as $M_{1}=10^{10}$ GeV or
$M_{1}=10^{13}$ GeV depending on the $U(1)_F$ charge for $N_1$. We have
found that the required inflaton masses and reheating temperature are
naturally realized in existing SUSY inflation models such as for
topological~\cite{Izawa-Kawasaki-Yanagida} or for 
hybrid inflation \cite{Hybrid-Inf-Model}.

%
\section*{Acknowledgements}
The author would like to thank T. Kanazawa
for useful discussion on hybrid inflation models.
This work is partially supported by the Japan Society for the
Promotion of Science (TA,KH) and ``Priority Area:
Supersymmetry and Unified Theory of Elementary Particles
($\sharp$707)''(MK,TY).



\clearpage
\baselineskip 0.5cm
\begin{thebibliography}{100}
\bibitem{Fukugita-Yanagida}
M. Fukugita and T. Yanagida,
Phys. Lett. {\bf B174} (1986) 45.
\bibitem{Kuzmin-Rubakov-Shaposhnikov}
V.A. Kuzmin, V.A. Rubakov, and M.E. Shaposhnikov,
Phys. Lett. {\bf B155} (1985) 36.
\bibitem{See-Saw}
T. Yanagida, 
{\it in} Proc. Workshop on the unified theory and 
the baryon number in the universe, (Tsukuba, 1979), 
{\it eds.} O. Sawada and S. Sugamoto,
Report KEK-79-18 (1979);\\
M. Gell-Mann, P. Ramond and R. Slansky,
{\it in} ``Supergravity''
(North-Holland, Amsterdam, 1979)
{\it eds.} D.Z. Freedman and P. van Nieuwenhuizen.
\bibitem{Super-K}
Y.~Fukuda {\it et al.} [Superkamiokande Collaboration],
Phys. Lett. {\bf B433} (1998) 9;
Phys. Lett. {\bf B436} (1998) 33;
Phys. Rev. Lett. {\bf 81} (1998) 1562.
\bibitem{LG-thermal}
See, for a recent review, 
W.~Buchm\"uller and M.~Pl\"umacher,
hep-ph/9904310;
M. Pl\"umacher,
Nucl. Phys. {\bf B530} (1998) 207 and
reference therein.
\bibitem{LG-infdecay}
B.A. Campbell, S. Davidson and K.A. Olive,
Nucl. Phys. {\bf B399} (1993) 111;
K.~Kumekawa, T.~Moroi and T.~Yanagida,
Prog. Theor. Phys. {\bf 92} (1994) 437;
G. Lazarides, hep-ph/9904428 and reference therein;
G.F. Giudice, M. Peloso, A. Riotto, I. Tkachev,
hep-ph/9905242.
\bibitem{LG-osc}
H.~Murayama and T.~Yanagida,
Phys. Lett. {\bf B322} (1994) 349 ;
H.~Murayama, H.~Suzuki, T.~Yanagida and J.~Yokoyama,
Phys. Rev. Lett. {\bf 70} (1993) 1912,
Phys. Rev. {\bf D50} (1994) 2356.
\bibitem{Izawa-Kawasaki-Yanagida} 
  K.I. Izawa, M. Kawasaki and T. Yanagida, 
  Prog. Theor. Phys. {\bf 101} (1999) 1129.
\bibitem{Hybrid-Inf-Model}
C.~Panagiotakopoulos,
Phys. Rev. {\bf D55} (1997) 7335;
A.~Linde and A.~Riotto,
Phys. Rev. {\bf D56} (1997) 1841.
\bibitem{Froggatt-Nielsen}
C.D.~Froggatt and H.B.~Nielsen,
Nucl. Phys. {\bf B147} (1979) 277.
\bibitem{Gravitino-Prob}
M.Y. Khlopov and A.D. Linde,
Phys. Lett. {\bf B138} (1984) 265;
J. Ellis, J.E. Kim and D.V. Nanopoulos,
Phys. Lett. {\bf B145} (1984) 181;
M. Kawasaki and T. Moroi,
Prog. Theor. Phys. {\bf 93} (1995) 879;
see also, for example, a recent analysis,
E. Holtmann, M. Kawasaki, K. Kohri and T. Moroi,
Phys. Rev. {\bf D60} (1999) 023506.
\bibitem{Murayama-Yanagida}
H.~Murayama and T.~Yanagida, in Ref. \cite{LG-osc}.
\bibitem{Covi}
L. Covi, E. Roulet and F. Vissani,
Phys. Lett. {\bf B384} (1996) 169;
M. Flanz, E.A. Paschos and U. Sarkar,
Phys. Lett. {\bf B345} (1995) 248;
Phys. Lett. {\bf B384} (1996) 487 (E).
\bibitem{Buchmuller-Plumacher}
W. Buchm\"uller and  M. Pl\"umacher,
Phys. Lett. {\bf B431} (1998) 354.
\bibitem{Buchmuller-Yanagida}
W.~Buchm\"uller and T.~Yanagida,
Phys. Lett. {\bf B445} (1999) 399.
\bibitem{Khlebnikov-Shaposhnikov}
S.Y. Khlebnikov and M.E. Shaposhnikov,
Nucl. Phys. {\bf B308} (1988) 885;
J.A. Harvey and M.S. Turner,
Phys. Rev. {\bf D42} (1990) 3344.
\bibitem{Sato-Yanagida}
J.~Sato and T.~Yanagida,
Talk given at 18th International Conference 
on Neutrino Physics and Astrophysics (NEUTRINO 98),
(Takayama, Japan, 1998),
hep-ph/9809307;
 P.~Ramond,
Talk given at 18th International Conference 
on Neutrino Physics and Astrophysics (NEUTRINO 98),
(Takayama, Japan, 1998),
hep-ph/9809401.
\bibitem{Topological-Inf} 
  A. Linde,
  Phys. Lett. {\bf B327} (1994) 208;
  A. Vilenkin,
  Phys. Rev. Lett. {\bf 72} (1994) 3137.
\bibitem{Sakai}
  N. Sakai, H. Shinkai, T. Tachizawa and K. Maeda,
  Phys. Rev. {\bf D53} (1996) 655;(E) Phys. Rev. {\bf D54} (1996) 2981.
\bibitem{Cho}
  I. Cho and A. Vilenkin,
  Phys. Rev. {\bf D56}  (1998) 7621;
  A.A. de Laix, M. Trodden and T. Vachaspati
  Phys. Rev. {\bf D57}  (1998) 7186.
\bibitem{COBE}
  C.L. Bennett et al.,
  Astrophys. J. {\bf 464} (1996) L1.
\bibitem{Lazarides}
G. Lazarides, in Ref. \cite{LG-infdecay}.
\bibitem{Dvali}
  G. Dvali, Q. Shafi and R.K. Shaefer,
  Phys. Rev. Lett. {\bf 73} (1994) 1886.
\bibitem{Copeland} 
  E.J. Copeland, A.R. Liddle, D.H. Lyth, E.D. Stewart 
  and D. Wands,
  Phys. Rev. {\bf D49} (1994) 6410.
\bibitem{String} 
  B. Allen, R.R. Caldwell, S. Dodelson, L. Knox, E.P.S. Shellard,
  and A. Stebbins,
  Phys. Rev. Lett. {\bf 79} (1997) 2624.
\bibitem{MAP} 
 http://map.gsfc.nasa.gov/;
 http://astro.estec.esa.nl/SA-general/Projects/Planck.
\end{thebibliography}
\end{document}